\documentclass{aa}
\usepackage{psfig,graphicx}
\begin{document}    

\thesaurus{06(08.03.4;08.05.2;08.09.2 4U\,0115+63;08.02.1;08.14.1;13.25.5)}

\title{The Be/X-ray transient 4U\,0115+63/V635 Cas}
\subtitle{II. Outburst mechanisms}

\author{I.~Negueruela\inst{1,2,3}
\and A.~T.~Okazaki\inst{4,5}
\and J.~Fabregat\inst{6}
\and M.~J.~Coe\inst{7}
\and U.~Munari\inst{8,9}
\and T.~Tomov\inst{8,9}}
                                                            
\institute{ Observatoire de Strasbourg, 11 rue de l'Universit\'{e}, F67000
Strasbourg, France
\and
SAX SDC, ASI, c/o Nuova Telespazio, via Corcolle 19, I00131
Rome, Italy
\and Astrophysics Research Institute, Liverpool John Moores University, 
Byrom St., Liverpool, L3 3AF, U.K.
\and Faculty of Engineering, Hokkai-Gakuen University,Toyohira-ku, Sapporo
062-8605, Japan
\and Institute of Astronomy, Madingley Road, Cambridge, CB3 0HA, U.K.
\and Departamento de Astronom\'{\i}a y Astrof\'{\i}sica, Universidad
de Valencia, 46100, Burjassot, Valencia, Spain
\and Physics and Astronomy Dpt., University of Southampton, Southampton,
SO17 BJ1, U.K.
\and Osservatori Astronomici di Padova e Asiago, via dell'Osservatorio 8,
  36012 Asiago (Vicenza), Italy
\and Centro Interdipartimentale di Studi ed Attivit\`{a} Spaziali (C.I.S.A.S.)
  "G.~Colombo", Universit\`{a} di Padova, Italy }

\mail{ignacio@isaac.u-strasbg.fr}

\date{Received    / Accepted     }

\titlerunning{Outburst mechanisms in 4U\,0115+63}
\authorrunning{Negueruela et al.}
\maketitle 

\begin{abstract}

We present multi-wavelength long-term monitoring observations of V635 Cas, 
the optical counterpart to the transient X-ray pulsar 4U\,0115+63. 
The evolution of emission lines and photometric magnitudes indicates
that the Be star undergoes relatively fast ($\sim 3-5 \: {\rm yr}$) 
quasi-cyclic activity, losing and reforming its circumstellar disc. We
show that the general optical, infrared and X-ray behaviour
can be explained by the dynamical evolution of the viscous circumstellar
disc around the Be star. After each disc-loss episode, the disc starts
reforming and grows until it reaches the radius at which the resonant
interaction of the neutron star truncates it. At some point,
the disc becomes unstable to (presumably radiative) warping
and then tilts and starts precessing. The tilting is very large and 
disc precession leads to a succession of single-peaked and shell profiles
in the emission lines. Type II X-ray outbursts take place after the disc has
been strongly disturbed and we speculate that the distortion of the disc
leads to interaction with the orbiting neutron star.
 We discuss the implications of 
these correlated optical/X-ray variations for the different models proposed 
to explain the occurrence of X-ray outbursts in Be/X-ray binaries.
We show that the hypothesis of mass ejection events as the cause of
the spectacular variability and X-ray outbursts is unlikely to be
meaningful for any Be/X-ray binary.
\end{abstract}

\keywords{stars: circumstellar matter -- emission-line, Be -- individual: 
4U\,0115+63, -- binaries:close -- neutron   -- X-ray: stars

\section{Introduction}

The hard X-ray transient 4U\,0115+63 (X\,0115+634) is one of the best 
studied Be/X-ray
binary systems (see Campana 1996; Negueruela et al. 1997, henceforth 
N97). This is the second on a series of papers dedicated to understanding
its behaviour and investigating what it can tell us about the general
class of Be/X-ray transients. An introduction to the source has been 
presented in Paper I (Negueruela \& Okazaki 2000), where it was
shown that V635 Cas is a B0.2Ve star at a distance of $\sim 7\,{\rm kpc}$.
The primary is seen under a moderate inclination ($40^{\circ}<i<60^{\circ}$),
and the high $v\sin i \sim 300\,{\rm km}\:{\rm s}^{-1}$ indicates that
it rotates close to its break-up velocity. A model for the system in 
which the neutron star orbits a $18\,M_{\sun}$ primary was adopted after
showing that the actual mass of the primary would have little impact on the
orbital parameters derived. Under those conditions, the 24.3-d eccentric
($e=0.34$) orbit results in periastron and apastron distances of 
$a_{{\rm per}} = 8\: R_{*}$ and $a_{{\rm ap}} = 16\: R_{*}$ respectively.

The disc around the Be star was modelled as a viscous decretion disc, i.e.,
a quasi-Keplerian disc held by the transport of angular momentum via
viscous interactions. The outflow (radial) velocity in such a disc is 
expected to be strongly subsonic, in agreement with all the observations
of Be stars in general and V635 Cas in particular. It was shown that 
such a disc cannot reach a steady state due to tidal and resonant 
interaction with the neutron star, and it is truncated at a radial
distance which depends on the value of the viscosity, but which for 
most reasonable values was found to correspond to the $4:1$ commensurability
of disc and binary orbital periods, which is situated at adistance of
0.39 times the semi-major axis. This truncation radius is closer to
the Be star than 
the first Lagrangian point $L_{1}$ for all orbital phases. Given that 
accretion of material from a very slow outflow implies that mass transfer
takes place through $L_{1}$, no material can reach the neutron star
under normal conditions, which explains the usual quiescence state of the
source, with complete absence of X-ray emission.

\section{Observations}

We present optical and infrared photometry and optical
spectroscopy of V635 Cas, taken with a large array of telescopes during
the last ten years. The observations cover a large variety of very
different states of the source.

\subsection{Optical spectroscopy}

Spectra of V635~Cas in the red region from a number of sources have been 
collected in this work. 
Data from 1990\,--\,1994 have been retrieved from the La Palma Archive
(Zuiderwijk et al. 1994) or taken from N97. These observations had 
been obtained with either
the 4.2-m William Herschel Telescope (WHT) or the  2.5-m Isaac Newton 
Telescope (INT), both located  at the Observatorio del Roque de los 
Muchachos, La Palma, Spain. Most spectra have been taken with the 
Intermediate Dispersion Spectrograph (IDS) on the INT, 
equipped with the 235-mm camera + R1200Y grating which gives a nominal 
dispersion of $\sim 0.8$ \AA/pixel with most of the cameras used. 
Higher resolution spectra were obtained with either IDS equipped with 
the 500-mm camera and the R1200Y grating or the Intermediate Dispersion 
Spectroscopic and Imaging System (ISIS) on the WHT, equipped with the 
R1200R grating. 

During 1995\,--\,1998 spectroscopy was regularly obtained with the INT,
equipped with the 235-mm camera + R1200Y grating and either the Tek 3
CCD (giving a nominal dispersion of $\sim 0.8$ \AA/pixel) or the 
EEV \#12 CCD (giving a nominal dispersion of $\sim 0.4$ \AA/pixel). 
Spectral resolutions (as estimated from the FWHM of
arc lines) varied according to slit width (generally matched to seeing)
from $\sim 1.2$ \AA\ to $\sim 2$ \AA.
All these data have been reduced using the {\em Starlink}
software packages {\sc ccdpack} \cite{dra98} and {\sc figaro} \cite{sho97} 
and analysed using {\sc figaro} and {\sc dipso} \cite{how97}.

\begin{table*}
 \caption{Observational details of the red spectroscopy and H$\alpha$
line parameters. Errors in the 
measurements of EW are typically 10 \%, due to the subjective continuum 
determination. See Section \ref{sec:shapes} for the interpretation of
the line parameters for different line shapes. When the date is followed
by a number, it indicates that the spectrum is shown in 
Figure~\ref{fig:cycle}.}
\begin{center}
\begin{tabular}{lccccccc}
\hline
Date & Telescope & Nominal & Shape &Total H$\alpha$&$\Delta v_{{\rm peak}}$  & FWHM & TBW\\
& & Dispersion (\AA/pix) & & EW (\AA) & (km s$^{-1}$) & (km s$^{-1}$)& (\AA)\\
\hline
Feb. 14, 1990 $^{(1)}$ & WHT & $\sim 0.8$ &SPV & $-$7.0 & $-$ & 580 & 23\\
Dec. 27, 1990 $^{(2)}$& INT & $\sim 0.4$ &SHS & $-$4.6 & 370 & 620 & 40 \\
Jan. 27, 1991 $^{(3)}$& INT & $\sim 0.8$ &SPS & $-$10.0 & $-$& 440 & 18 \\
Aug. 28, 1991 $^{(4)}$& INT & $\sim 0.4$ &SHV & $-$4.3 & 420 & 635 & 52 \\
Dec. 14, 1991 $^{(5)}$& INT & $\sim 0.4$ &DPV & $-$5.4 & 480 & $-$ & 22 \\
Aug. 05, 1992 $^{(6)}$ & INT & $\sim 0.8$ &DPR & $-$4.1 & 425 & 705 & 47\\
Jan. 01, 1993 & WHT & $\sim 0.4$ &DPV & $-$7.0 & 390 & 730 & 42\\
Sep. 23, 1993 & PAL & $\sim 0.8$ &DPR & $-$4.9 & 410 & $-$ & 44\\
Dec. 18, 1993 & WHT & $\sim 0.4$ &DPS & $-$5.5 & 405 & 680 & 48\\
Dec. 19, 1993 $^{(7)}$& WHT & $\sim 0.4$ &DPS & $-$5.3 & 390 & 670 & 44\\
Jan. 09, 1995 $^{(8)}$ & INT & $\sim 0.8$ &SPV & $-$8.8 & $-$ & 455 & 25\\
Jul. 03, 1995 & INT & $\sim 0.8$ &SHV & $-$3.6 & 390 & 635 & 55\\
Sep. 12, 1995 $^{(9)}$& INT & $\sim 0.4$ &SHV & $-$5.0 & 420 & 600 & 55\\
Oct. 15, 1995 & Asiago & $\sim 7.8$& SP? & $-$10.2 & $-$ & $-$ & $-$\\
Nov. 29, 1995 & JKT & $\sim 1.2$ &SPR & $-$10.5 & $-$ & $-$ & $-$\\
Jan. 12, 1996 & INT & $\sim 0.8$ &DPR & $-$11.3 & 210 & 425 & 50 \\
Jan. 31, 1996 & INT & $\sim 1.5$ &DPR & $-$8.0 & 280 & 560 & 47\\
Jun. 20, 1996 $^{(10)}$& INT & $\sim 0.8$ &DPR & $-$7.4 & 340 & 590 & 54\\
Jul. 09, 1996 & INT & $\sim 0.8$ &DPR & $-$6.5 & 390 & $-$ & 53\\
Aug. 26, 1996 & INT & $\sim 0.8$ &SPR & $-$8.4 & $-$ & 550 & 24\\
Feb. 01, 1997 $^{(11)}$& INT & $\sim 0.8$ &DPR & $-$3.6 & 480 & $-$ & 27\\
Jul. 17, 1997 $^{(12)}$& WHT & $\sim 0.4$ &ABS & $+$1.3 & $-$ & 460 & $-$ \\
Jul. 25, 1997 & INT & $\sim 0.8$ &ABS & $+$1.2 & $-$ & 430 & $-$\\
Aug. 23, 1997 & INT & $\sim 0.8$ &ABS & $+$1.1 & $-$ & 430 & $-$\\
Sep. 05, 1997 & INT & $\sim 0.8$ &ABS & $+$1.1 & $-$ & 350 & $-$\\
Sep. 25, 1997 & INT & $\sim 0.8$ &ABS & $+$1.0 & $-$ & 480 & $-$\\
Oct. 01, 1997 & INT & $\sim 0.8$ &ABS & $+$1.1 & $-$ & 330 & $-$\\
Oct. 13, 1997 & INT & $\sim 0.8$ &DPR & $+$0.2 & 700 & $-$ & 23\\
Nov. 14, 1997 $^{(13)}$& WHT & $\sim 0.4$ &DPS & $-$0.4 & 610 & $-$ & 24\\
Aug. 02, 1998 $^{(14)}$& INT & $\sim 0.4$ &DPS & $-$6.5 & 305 & 700 &50\\
Aug. 07, 1998 & INT & $\sim 0.4$ &DPV & $-$6.5 & 325 & 660 &50\\
Sep. 10, 1998 & INT & $\sim 0.8$ &DPV & $-$7.4 & 310 & 690 &51\\
Sep. 25, 1998 $^{(15)}$& INT & $\sim 0.4$ &DPV & $-$7.5 & 280 & 610 &46\\
Dec. 28, 1998 $^{(16)}$& INT & $\sim 0.8$ &SPR & $-$8.7 & $-$ & 550 & 26\\
\hline
\end{tabular}
\end{center}
 \label{tab:halpha}
\end{table*}

Some low-resolution spectra have been secured with
the Boller \& Chivens spectrograph attached to the 1.82-m telescope
operated by the Osservatorio Astronomico di Padova atop of Mount Ekar,
Asiago (Italy). The detector has been a Thompson TH7882 UV-coated CCD,
580x388 pixels of 23$\mu$m size. We used a 150 ln/mm grating giving a
dispersion of 7.8 \AA/pixel. The slit was set to 1.5 arcsec,
for a PSF on the spectrograph focal plane of $\sim$2.5 pixels. One
spectrum was taken with the 1.5-m telescope at 
Palomar Mountain Observatory (PAL) and one spectrum was taken with
the 1-m Jakob  Kapteyn Telescope (JKT) at La Palma.

A complete log of H$\alpha$ spectroscopy is presented in Table 
\ref{tab:halpha}. Low-resolution spectra of the region $\lambda\lambda
6800\,-\,9500$ \AA\ were taken on October 22, 1994 (Asiago) and August 12,
1998 (WHT). In both cases, no obvious stellar features are visible, though
the 1998 spectrum could show Pa13 and Pa12 in emission. A 
higher resolution spectrum taken on September 10, 1998, with the INT
shows that the upper Paschen lines have weak emission components that
fill in the photospheric absorption features, resulting in the 
featureless continuum seen at lower resolutions.

\subsection{Optical photometry}

Several sets of $UBVRI$ photometry have been obtained using the
Jacobus Kapteyn Telescope (JKT), located  at the Observatorio del Roque 
de los Muchachos, La Palma, Spain. The telescope was equipped with the 
Tek4 CCD and the Harris filter set. 
Instrumental magnitudes were extracted through synthetic aperture routines
contained in the IRAF package. Transformation to the Johnson/Cousins
system was made by means of a system of secondary standard stars on the
same CCD frame\footnotemark \footnotetext{Finding charts containing the 
local photometric sequence can be found at the following World Wide Web 
address: \textsf{http://www.soton.astro.ac.uk/$\sim$ind/v635cas.html}}, 
calibrated through observations of photometric standards
from Landolt (1992) in an earlier photometric run (N97),
except for the October 11th 1997 observation, which was part of a larger 
observing programme and was transformed through calibrations 
derived from observations of a number of Landolt (1992) standard stars taken 
on the same night. The photometric errors are the estimated uncertainties 
in the transformation equations. The data are listed together with the 
observations from N97 in Table~\ref{tab:optphot}

\begin{table*}
\caption{Observational details of the optical photometry. All the
observations are from the JKT. The first three datasets are taken from N97.}
\begin{center}
\begin{tabular}{llllllc}
\hline
Date & $U$ & $B$   &  $V$ &   $R$ &   $I$ & $B-V$ \\
\hline
Dec. 03, 1993 & &16.90 & 15.46 & 14.50 & 13.29 & 1.44 \\
Aug. 18, 1994 & &16.30 & 14.92 & 13.95 & 12.75 & 1.38 \\
Jan. 19, 1995 & &16.40 & 14.79 & 13.68 & 12.46 & 1.61 \\
\hline
Oct. 11, 1997 &17.07$\pm$0.10 & 16.86$\pm$0.03 & 15.41$\pm$0.02 & 14.52$\pm$0.03 &13.17$\pm$0.08 & 1.45 \\
Jan. 07, 1998 & 17.05$\pm$0.10& 16.92$\pm$0.07 & 15.53$\pm$0.04 & 14.55$\pm$0.03 & 13.49$\pm$0.03 & 1.39\\
Sep. 05, 1998 & 17.04$\pm$0.08 &16.93$\pm$0.06 & 15.37$\pm$0.02 & 14.32$\pm$0.02 & 13.19$\pm$0.02 & 1.56 \\
\hline
\end{tabular}
\end{center}
 \label{tab:optphot}
\end{table*}

Photographic photometry has been obtained over 1987\,--\,1997 with the
67/92-cm Schmidt telescope operated by Osservatorio Astronomico di Padova
atop Mt. Ekar, Asiago (Italy). The telescope has a 205-cm focal length
and UBK7 corrective plate. For the $B$ band the standard emulsion/filter
combination 103a-O + GG13 has been adopted. The brightness of V635 Cas
has been estimated at the microscope against the same comparison sequence
used for CCD photometry. The results of photographic photometry are reported
in Table~\ref{tab:bplates}.

\begin{table}
\caption{Estimates of the $B$ magnitude of V635 Cas based on plates
taken with the 67/92-cm Schmidt telescope of Asiago Observatory. The
last column is an estimate of the quality of the plate ranging from 
A (very good) to D (poor).}
\begin{center}
\begin{tabular}{cc|r|l|c}
\hline
plate \# &date&lim.mag& $B$ &quality\\ 
\hline
 13792 & 29-04-87 & 19.0 & 16.3  & D   \\
 13834 & 01-07-87 & $<$19.2 & 16.6  & B   \\
 13850 & 21-07-87 & $<$19.2 & 16.8 & B   \\
 13938 & 21-11-87 & 19.4 & 16.8  & A/B \\
 13972 & 19-12-87 & $<$19.2 & 16.3 & C   \\
 14002 & 11-02-88 & 17.4 & 16.4 & B   \\
 14642 & 07-10-89 & 17.4 & 16.6  & B   \\
 14702 & 26-11-89 & 17.4 & 16.6  & C/B \\
 14924 & 13-11-90 & 17.7 & 17.1  & B   \\
 14979 & 18-01-91 & 17.4 & 16.3  & B   \\
 15002 & 14-02-91 & 17.1 & 16.3  & A   \\
 15353 & 26-12-92 & $<$19.2 & 16.6 & B/C \\
 15387 & 17-01-93 & $>$19.2 & 16.7  & B   \\
 15395 & 18-01-93 & $>$19.2 & 16.6  & C   \\
 15405 & 19-01-93 & $>$19.2 & 16.5  & B   \\
 15411 & 21-01-93 & 19.3 & 16.5 & B/C \\
 15416 & 22-01-93 & 19.3 & 16.4  & B   \\
 15423 & 24-01-93 & 19.3 & 16.5  & B   \\
 15427 & 25-01-93 & 19.2 & 16.7  & C   \\
 15434 & 27-01-93 & $>$19.2 & 16.5  & C   \\
 15438 & 13-02-93 & $>$19.2 & 16.6  & D   \\  
 15448 & 14-02-93 & $<$19.2 & 16.7  & B   \\
 15656 & 15-10-93 & $<$19.2 & 16.8  & B/A \\
 15706 & 16-12-93 & 19.4 & 16.9 & B   \\
 15721 & 15-01-94 & $<$19.2 & 16.8  & B/C \\
 15804 & 10-10-94 & $>$19.2 & 16.7 & B   \\
 16223 & 12-01-97 & $<$19.2 & 16.8  & B   \\ 
\hline
\end{tabular}
\end{center}
 \label{tab:bplates}
\end{table}

\begin{figure*}
\begin{picture}(500,600)
\put(0,0){\includegraphics{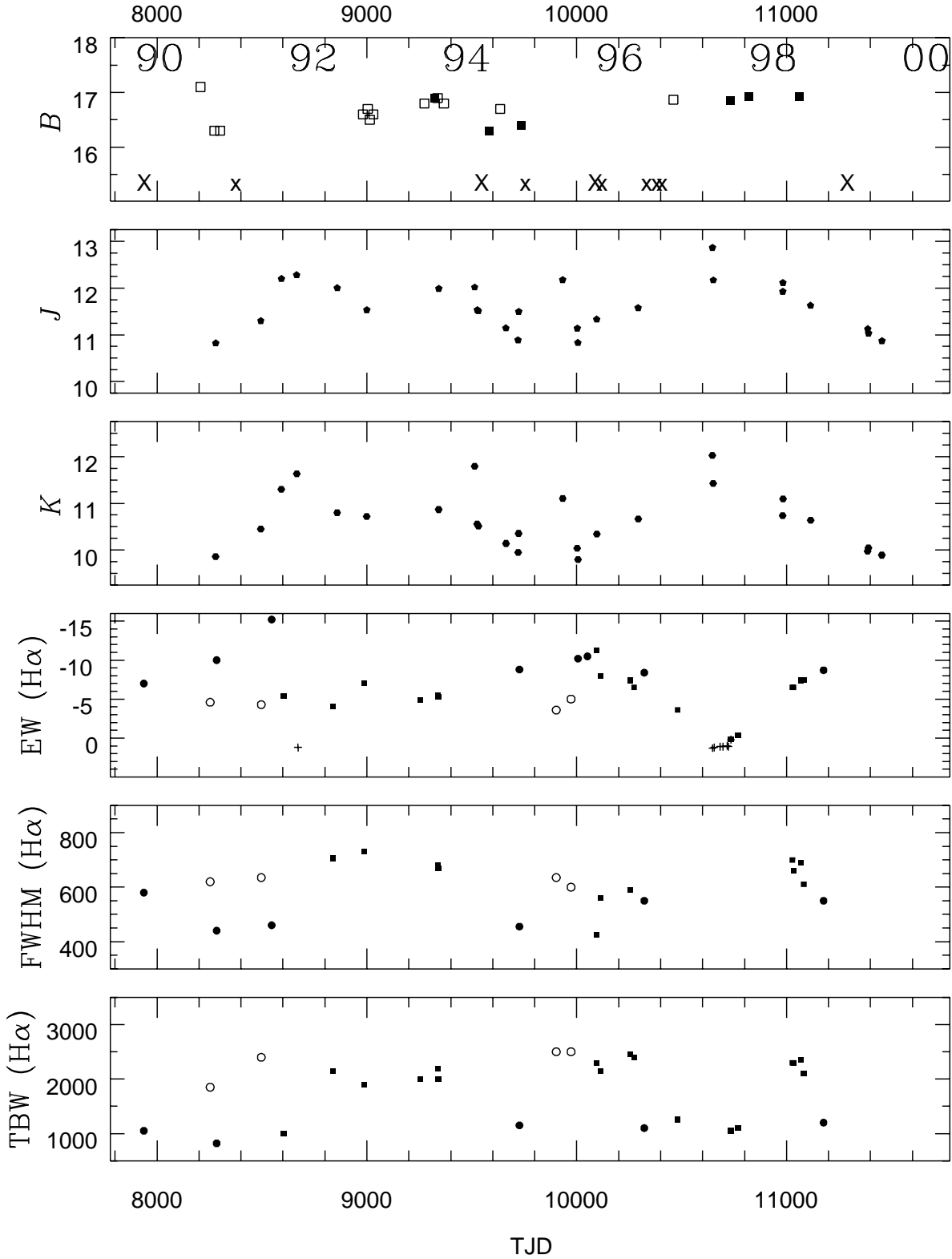}}
\end{picture}
\caption{A general view to the evolution of the infrared and optical 
lightcurves and H$\alpha$ line parameters for V635 Cas during 1990\,--\,1998. 
In the top panel, $B$ magnitudes are represented 
by squares (open when they are plate estimates and filled when they
are the result of CCD photometry). Times of X-ray activity are marked.
Capital 'X' indicates a Type II outbursts, while 'x' marks a Type I outburst
or flare. In the three bottom panels, displaying the H$\alpha$ line 
parameters, absorption lines are represented by crosses, single-peaked
lines are shown as filled circles, shell profiles are open circles
and double-peaked lines are filled squares (the FWHM and TBW of
absorption lines is not plotted, since it has a completely different
physical meaning; FWHM is not plotted either for some profiles so 
asymmetric that it loses any meaning). The two cycles of disc
loss and reformation seen in the emission lines are clearly mirrored
in the evolution of the $J$ and $K$ magnitudes, but do not appear 
clearly reflected in the $B$ lightcurve or in the evolution of the 
$(J-K)$ colours.
Error bars have been removed for clarity 
(see Tables for details)}
\label{fig:allphot}
\end{figure*}

\subsection{Infrared Photometry}
 
The source has been monitored in the infrared since 1991 using the 
Continuously Variable Filter
(CVF) on the 1.5-m Carlos Sanchez Telescope (TCS) at the Teide
Observatory, Tenerife, Spain. Many of the observations have already 
been reported in N97 and Negueruela et al. (1998, henceforth N98). 
However, we have 
reprocessed the complete dataset using the procedure described by Manfroid
(1993) and all instrumental values have been transformed to the TCS standard 
system (Alonso et al. 1998) in order to generate a more coherent dataset.

\begin{table*}
\caption{Observational details of the infrared photometry. All the
observations are from the TCS.}
\begin{center}
\begin{tabular}{lcccc}
\hline
Date & TJD & $J$ mag  &  $H$ mag &   $K$ mag\\
\hline
Jul 27, 1996&10292.7& 11.58$\pm$0.04&11.10$\pm$0.01&10.67$\pm$0.02\\
Jul 28, 1996&10293.7& 11.68$\pm$0.04&11.08$\pm$0.01&10.72$\pm$0.03\\
Jul 29, 1996&10294.6& 11.55$\pm$0.05&11.10$\pm$0.03&10.72$\pm$0.02\\
Jul 17, 1997&10646.7& 12.87$\pm$0.16&12.15$\pm$0.13&12.03$\pm$0.14\\
Jul 18, 1997&10648.7& 12.67$\pm$0.09&12.06$\pm$0.05&11.75$\pm$0.04\\
Jul 19, 1997&10649.7& 12.45$\pm$0.05&11.93$\pm$0.04&11.77$\pm$0.03\\
Jul 20, 1997&10650.7& 12.17$\pm$0.05&11.67$\pm$0.04&11.43$\pm$0.04\\
Jul 21, 1997&10651.7& 12.46$\pm$0.05&11.89$\pm$0.04&11.72$\pm$0.06\\
Jun 16, 1998&10981.7& 11.93$\pm$0.04&11.26$\pm$0.03&10.73$\pm$0.05\\
Jun 17, 1998&10982.7& 12.11$\pm$0.06&11.56$\pm$0.02&11.10$\pm$0.02\\
Oct 27, 1998&11114.4& 11.63$\pm$0.07&11.17$\pm$0.04&10.64$\pm$0.06\\
\hline
\end{tabular}
\end{center}
 \label{tab:irphot}
\end{table*}

Since the new values do not differ significantly from those published
when compared to the range of variability of the source, in 
Table~\ref{tab:irphot} we only report the data for the period 1996\,--\,1998.
Measurements for 1991\,--\,1995, which are also shown in 
Fig.~\ref{fig:allphot}, can be found in N97 and N98.

\begin{figure*}[ht]
\begin{picture}(500,350)
\put(0,0){\includegraphics{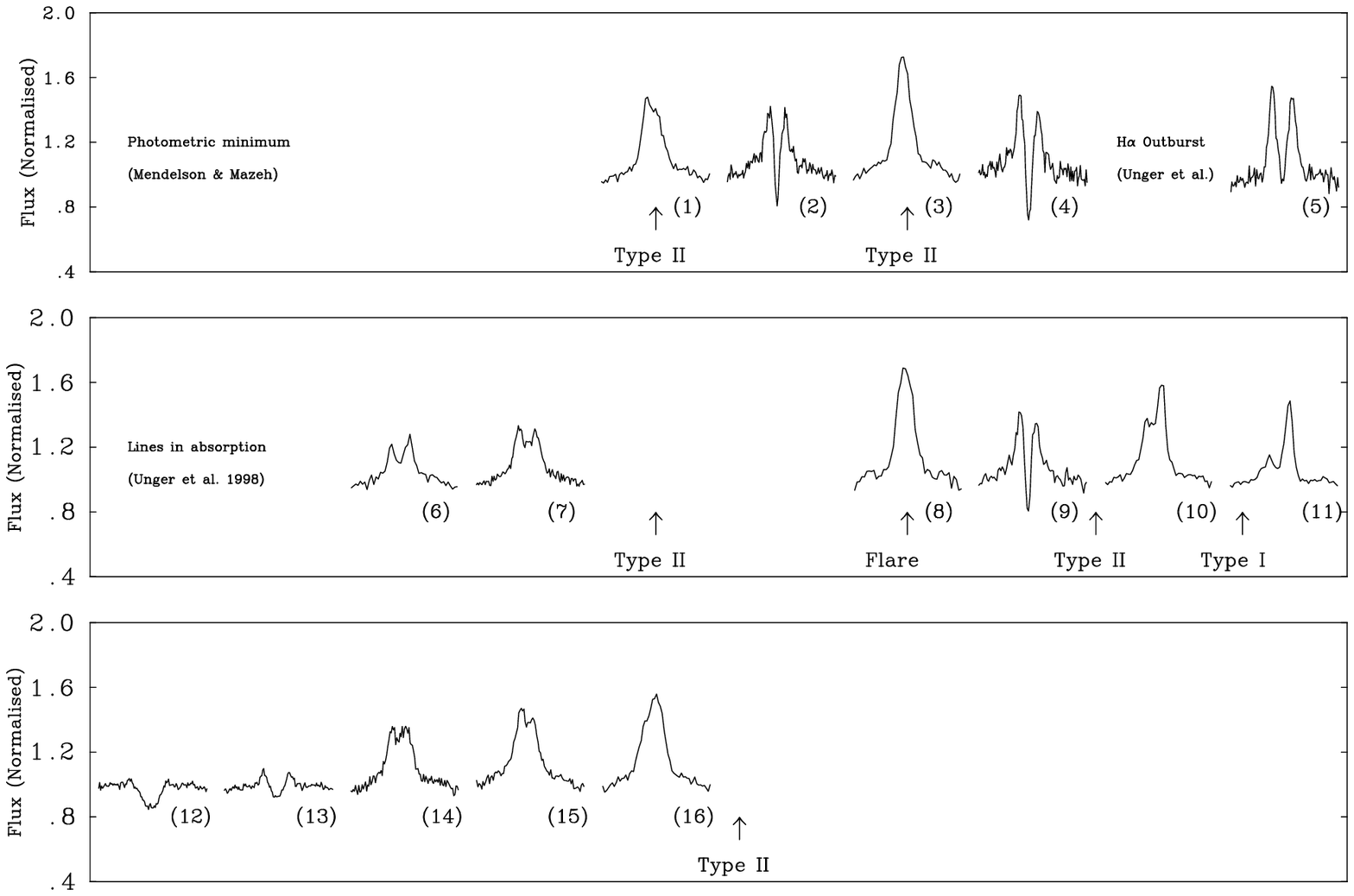}}
\end{picture}
\caption{The three panels show representative H$\alpha$ lines from
the three last cycles. Unfortunately the coverage is not complete. 
Every cycle starts with a disc-less state and ends in the dissipation
of the disc, which leads to the next disc-less state. The disc reappears
with some asymmetry already present (density wave?). As the strength
of the line increases, the two peaks converge and the line width
diminishes (disc growth). This is followed by a number of fast 
transitions between single-peaked and shell profiles (precession).
Finally, the strength of the line decreases and the disc disappears
(dissipation). Note that all spectra have been divided by a polynomial 
fit to the continuum and normalised to the same scale.
The numbers refer to Table \ref{tab:halpha}. The times at which X-ray 
outbursts occurred are marked, but note that the horizontal (temporal)
scale is not linear.}
\label{fig:cycle}
\end{figure*}

\section{Results}

4U\,0115+63 is the most active of Be/X-ray transients. Since 1969, it has
been observed to undergo thirteen Type II outbursts (6 before the discovery
of the optical counterpart) and several smaller 
flares (see Campana 1996). During the period covered by
our observations, five outbursts have taken place: in 
February 1990 (Tamura et al. 1992), March and April 1991 
(double-peaked outburst according to Cominsky et al. 1994), 
May 1994 (N97), November 1995 (Finger et al. 1995)
and March 1999 (Heindl et al. 1999). Smaller flares took 
place in January 1995 and January
1996 (Scott et al. 1996) and a series of four Type I outbursts
occurred between August and November 1996 (Bildsten et al. 1997; N98)

Between 1990 and early 1995, our optical observations are relatively sparse. 
The gaps
in our spectroscopic sampling are typically larger than the time-scale
for large changes in the shape of the lines and therefore we cannot derive
any firm conclusions on the correlation between the spectroscopic behaviour
and the X-ray activity. However, it is noteworthy that we did obtain
spectroscopy of the source on or close to 
three of the occasions when 
it was active in the X-rays (February 1990, January 1991 and January 1995)
and observed the emission line to be single-peaked, while all observations
during quiescence showed double-peaked (sometimes shell-like) weaker lines.

From the summer of 1995 our spectroscopic coverage has been much more
complete -- though still presenting some large gaps. This has allowed us
to study in detail the changes in the emission lines that have taken
place during 1995\,--\,1997, which are presented in the following
subsections.

\subsection{Disc loss and reformation}
\label{sec:disc}

Among the many different kinds of variability that Be stars display,
the most obvious is the transition between Be phases and normal B
phases. These changes have been observed in a large number of Be stars,
with and without companions, and constitute one of the defining 
characteristics of the Be phenomenon.  The absorption spectrum seen
is typical of a normal B star and the transitions are interpreted 
as due to the loss of the circumstellar disc.

Our observations (see Fig. \ref{fig:discloss}) reveal that during 1996
the strength of the emission lines in V635~Cas gradually declined 
and sometime between February and
July 1997 (unfortunately the period during which the source is too low
to be observed from La Palma), the absorption lines
typical of the photosphere of an early-type star became visible.
Similar disc-less have been reported in the
Be/X-ray binaries X Persei (Roche et al. 1997) and V725 Tau/A\,0535+26
(Haigh et al. 1999). The absorption spectrum seen is typical of a normal 
B star and the transitions are interpreted as due to the loss of the
circumstellar disc.

\begin{figure}[ht]
\begin{picture}(250,320)
\put(0,0){\includegraphics{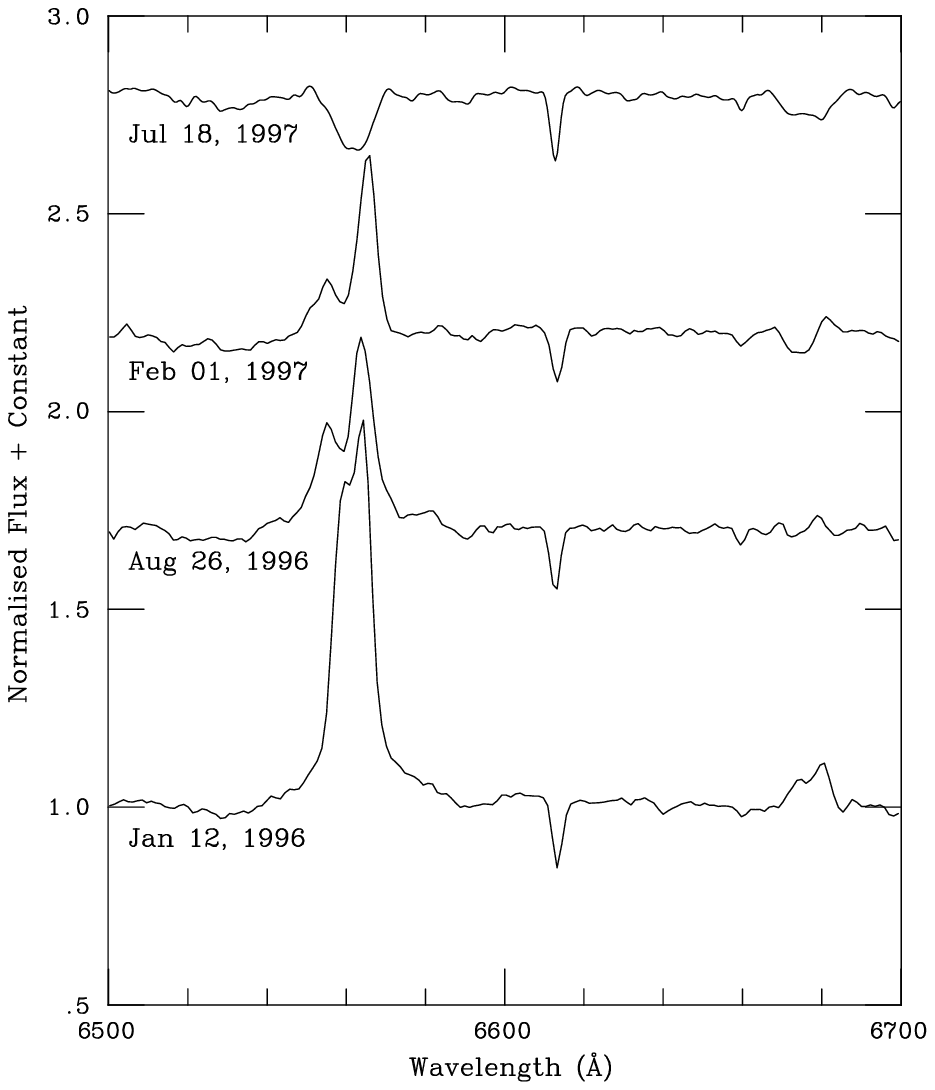}}
\end{picture}
\caption{Series of red spectra of V635 Cas showing the gradual weakening
and final disappearance of the emission lines during 1996 and
early 1997. Note the increase in the asymmetry as the line strength
decreases. All spectra have been 
divided by a spline fit to the continuum for normalisation and arbitrarily 
offset for clarity.
 See Table \ref{tab:halpha} for observational details of the spectra.}
\label{fig:discloss}
\end{figure}

Several spectra taken during the summer and autumn of 1997, following the 
dispersion of the envelope (see examples in Fig. 2 of Paper I) show 
the profile of H$\alpha$ basically reflecting 
the photospheric absorption feature, though some residual emission 
seems to be present in all the spectra (the effect is much stronger
in the \ion{He}{i}~$\lambda\lambda$~6678, 7065~\AA\ lines).
In the case of the disc-less phase of the Be/X-ray binary X Persei, 
Telting et al. (1998) conclude that even though the low photometric 
state lasted many months, the photospheric lines were seen without any
contamination for only $\sim 4$ weeks. After this, weak emission was
observed in the line wings. Therefore it is likely that in the case of 
V635~Cas there was a short time-span during which the lines were purely
photospheric, but this must have taken place as soon as the lines
reverted to absorption, i.e., during the gap in our observations.
Stronger highly-variable emission wings are seen in the spectra from
October and November 1997, as is typical of Be stars showing
low-level activity during an extended disc-less state (see Hanuschik 
et al. 1993; Rivinius et al. 1998 for examples)

The presence of absorption lines coincides with a photometric low state,
as can be readily seen by comparing the infrared magnitudes during July
1997 to those of July 1996 and June 1998. This is expected, since free-free
emission from the circumstellar disc contributes significantly to the
photometric magnitudes (N97). However, the very large variability observed
in the infrared magnitudes during one week in July 1997 shows that even
when the disc is basically absent, low-level activity still results
in the presence of some circumstellar material that contributes significantly
to the total brightness. Because of this, it is unlikely that any single
photometric dataset can be thought to characterise a ``state'' of the 
source. It is significant that the $BVRI$ measurements from October 1997 are 
very similar to those taken at other epochs when the disc was present, while
those from January 1998 show a decrease in $I$ by $0.32 \:\: {\rm mag}$
with respect to the October 1997 values.
Since the January 1998 observation presents the faintest and bluest
magnitudes in our dataset and the $I$-band magnitude (which is
likelier to suffer from circumstellar contamination than bluer magnitudes)
measured on that date is as faint as
the faintest point measured by Mendelson \& Mazeh (1991, henceforth MM91) 
in the five years 
spanned by their photometric monitoring of the source, we expect
these magnitudes to be close to the
actual apparent magnitudes of the star without any contribution from a disc.

This is consistent with the fact that,
even though all visual and infrared magnitudes of V635 Cas fluctuate 
continuously, the $V$ magnitude
has a tendency to take a value of $V \approx 15.5$ when the source is faint. 
The $B$-band light-curve presents smaller fluctuations than the $V$ band and, 
again, the values during faint states seem to be consistently $B=16.9$ . 
In Paper I, we have shown that the colours derived from the January 1998
observations are consistent with the intrinsic colours of a B0V star with 
the measured interstellar reddening and consequently 
used this photometric dataset to determine the distance to the source. 
Therefore we conclude that, in spite of the observed activity
during October and November 1997, the source was basically disc-less in
January 1998.

After the disc-less state, disc reformation was very quick. As can be 
seen in Table~\ref{tab:halpha} and Fig~\ref{fig:warp}, typical emission 
lines were present by August 1998, 
only $\sim 6$ months after the photometric minimum.
We note that the lines had been observed to be in absorption in February
1992 (Unger et al. 1998). Our observations show that emission was again
present in August 1992. Also in this case, the time for disc reformation is 
constrained to be $\la 6$ months.

In both cases, the line profile after the disc-less phase was double-peaked
and relatively broad, as corresponds to a Be star seen under a moderately
large inclination (Hummel 1994; Hanuschik et al. 1996). In Paper I, a 
moderate inclination of $40^{\circ}<i<60^{\circ}$ was derived. 
Therefore this particular line profile must be 
interpreted as the ``quiescence'' shape for V635 Cas, in the sense
that it reflects the inclination of the system and corresponds to an
unperturbed quasi-Keplerian disc (in the sense defined by Hanuschik 1996), 
while other line profiles correspond
to perturbed states of the disc.

\subsection{Timescales}
\label{sec:time}

Whitlock et al. (1989) were the first to notice that the X-ray activity 
of 4U\,0115+63 seemed to occur
with a quasi-periodicity of $\sim 3\:{\rm yr}$ during 1974\,--\,1989.
This quasi-periodicity was broken when the February 1990 outburst was
followed by another outburst in March 1991. However, the gap between this
outburst and the following one, in May 1994, was of slightly over three
years. During the 1992-1997 cycle, there were two major outbursts and 
several Type I outbursts and flares, but the gap between the 
November-December 1995 Type II outburst and the next Type II
outburst in March 1999 was again $\sim 3$ years.

Our observations reveal that during the last two 3-year gaps the
circumstellar disc dispersed and reformed. If the previous
photometric low state in 1989 (MM91) was also due to a disc loss episode, 
this would mean that the typical gap of $\sim 3$ years between X-ray
outbursts and the $\sim 3$-year cycles of disc loss and reformation 
occur in phase. Given that the X-ray source is powered by accretion 
of material from the circumstellar disc, we conclude that the X-ray 
quasi-periodicity is induced by the typical time-scales for disc 
loss and reformation.

\begin{figure}[ht]
\begin{picture}(250,320)
\put(0,0){\includegraphics{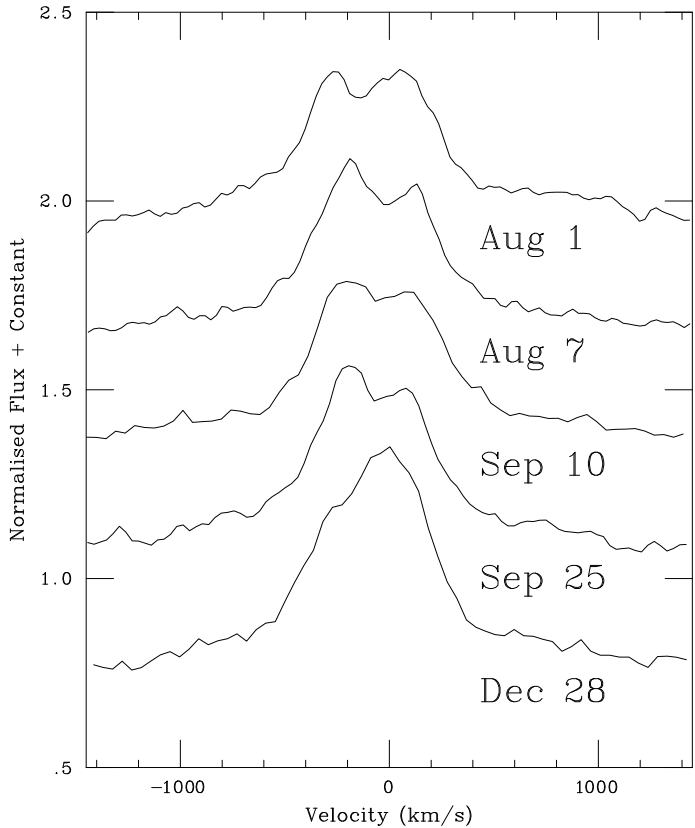}}
\end{picture}
\caption{Evolution of the H$\alpha$ line during 1998. Compare this series
with that shown on Fig.~\ref{fig:vrvar} (for the 1992\,--\,1993 period). 
As the line strength grows, both the peak separation and FWHM
decrease, indicating a growing disc. 
All spectra have been 
divided by a spline fit to the continuum for normalisation and arbitrarily 
offset for display. See Table~\ref{tab:halpha} for observational
details of the spectra.}
\label{fig:warp}
\end{figure}

Since all the variability in the line profiles, infrared magnitudes and
X-ray activity seems to be correlated and driven by this disc 
growth/dispersion cycle (see Fig.~\ref{fig:allphot} where the cyclicity
is apparent in the infrared magnitudes), in what follows we will
divide the time covered by these observations in:

\begin{itemize}
\item{\bf Cycle A: } Starting with the low photometric state in 1989
(which we interpret as a disc-loss event) and ending at the disc-less
episode in early 1992. Two X-ray outbursts were observed in February 1990
and March/April 1991.

\item{\bf Cycle B:} Starting with the disc-less state in 1992 and lasting
till the disc-less state in 1997. Two Type II outbursts took place in
May 1994 and November 1995, as well as some smaller flares.

\item{\bf Cycle C:} Starting with the disc-less state in 1997 and 
still continuing. A Type II X-ray outburst took place in March 1999.

\end{itemize}

A second time-scale observed is that of line changes. We find that the
typical time-scale for large changes is $\sim 1$ month (the shell to 
single-peaked transitions observed in January 1991 and October 1995) 
while a particular
line shape seems to exist for several months. 
In the following sections, we will discuss these line changes in detail
and conclude that they likely arise from the precession of the warped
disc of V635~Cas.

A third, less obvious time-scale could characterise the X-ray activity
during each of the cycles. During cycle B, X-ray coverage of the source 
has been continuous. The source has been active in four 
occasions. There were two large outbursts in May 1994 and November 1995,
a small flare in January 1995 and a series of small (Type I) outbursts
after August 1996. We note that the time interval between this four
periods of activity (May 1994\,--\,Jan 1995\,--\,Nov 1995\,--\,Aug 1996)
has always been $\sim$ 8\,--\,9 months. Such a separation is comparable 
to the gap between the two outbursts seen in cycle A
(though coverage was not complete around this time and smaller flares could
have gone undetected) and the $\sim 6$ month quasi-period observed between
four weak outbursts that took place during 1969\,--\,1970 
(Whitlock et al. 1989).
This third time-scale is also probably associated with the 
dynamics of the circumstellar disc, as discussed in the
next sections.

\subsection{Disc warping}
\label{sec:shapes}

As can be seen in Fig.~\ref{fig:cycle}, the H$\alpha$ emission line is 
extremely variable. The changes affect not only its shape but also
the Full Width at Half Maximum (FWHM) and in the Total Base Width
(TBW). Other emission lines, like \ion{He}{i}~$\lambda$6678\AA, undergo
a parallel evolution (see Fig.~\ref{fig:discloss} and also N98).
In Table \ref{tab:halpha}, the shape of H$\alpha$ in every 
spectrum has been indicated according to the following convention: 
when the line is fundamentally photospheric absorption it is labelled
ABS; emission lines have been marked as single-peaked (SP), double-peaked
(DP) or shell (SH), followed by a letter that describes the symmetry
of the profiles, i.e., red dominated (R), blue dominated (V) or 
symmetric (S) when the two peaks have a similar strength or the single
peak looks symmetric. Thus, SPR means a single peaked profile with
the peak on the red side of the line profile.

The different shapes of emission lines seen in different Be stars 
are in principle attributed to the different inclinations with respect 
to the line of sight of an equatorial circumstellar disc (see 
Hanuschik et al. 1995; Hummel 1994). For low
inclination angles, single-peaked profiles are seen (generally showing 
flank inflections due to non-coherent scattering, producing the wine-bottle
profile). For intermediate inclinations, double-peaked profiles are
seen due to Doppler broadening. For large inclinations, the outer cooler
regions of the disc intercept the line of sight and give rise to
shell profiles, characterised by deep narrow absorption cores that go
below the continuum level, at least in optically thin lines.

\begin{figure}[ht]
\psfig{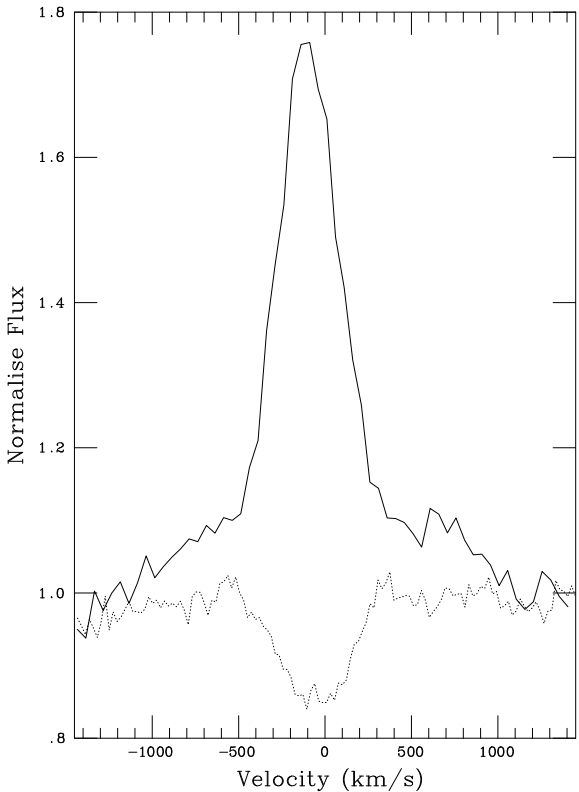} 
\caption{The complex shape of the H$\alpha$ emission line. The spectrum from
27th Jan 1991 is compared to the absorption spectrum observed in July 1997.
Note the broad emission component forming a base to the main feature.
In Table~\ref{tab:halpha}, the TBW of the single-peaked profiles represents
the width of the main feature and does not include this broad feature.
The two spectra have been divided by 2nd degree polynomial fits to the local
continuum.}
\label{fig:tbw}
\end{figure}

Hanuschik (1996) proposed that, for a restricted interval of 
inclinations around $i \simeq 70\degr$, changes
in the disc radius or thickness can cause the line shapes to change
between double-peaked and shell, but this cannot be the 
reason for the shell episodes of V635 Cas, since it also displays 
single-peaked profiles. As a matter of fact, shell profiles were only
observed as part of fast shell/single-peak transitions that took
place during the last phases of cycles A and B, leading in both cases
to the dispersal of the disc. Similar phenomenology (generally referred to
as spectacular variations) in the isolated Be stars $\gamma$~Cas and
59~Cyg was explained by Hummel (1998) in terms of variations in 
the angle under which
we are seeing the circumstellar disc due to its precession after it
has tilted off the equatorial plane.

In this picture the circumstellar disc evolves from an intermediate 
line-of-sight inclination (corresponding to the equatorial plane of 
the Be star) to a much lower inclination angle. When the disc
is basically seen face-on, we see single-peaked lines and the 
associated brightening because we observe emission from the 
whole disc surface. As the disc precesses we come to see it 
under a large inclination angle and its external regions
intercept the light from the inner disc, resulting in the observation
of shell lines and a faint state due to 
self-absorption of disc and
stellar emission.

Our dataset is far from having the time resolution of the $\gamma$
Cas observations that were used to derive this interpretation, and 
provides only a few snapshots of the evolution of line shapes. We are
not able to precisely identify the occurrence of full single-peaked
or shell episodes and therefore when considering a spectrum we cannot
in general decide to which phase of a shell/single-peak transition
it corresponds. However, we note the following similarities:

\begin{enumerate}
\renewcommand{\theenumi}{(\Roman{enumi})}

\item As $\gamma$ Cas, V635~Cas is seen under a moderate inclination angle
and, when an unperturbed disc is present, it displays double-peaked
symmetrical profiles.

\item As in the case of $\gamma$ Cas, the shell/single-peak transition 
events end up in the dispersion of the circumstellar disc. This happens
in both cycle A and B.

\item As in the case of $\gamma$ Cas, V/R variability occurs during the
transitions in both cycle A and B. In cycle B, we know that it started 
before the onset
of single-peak/shell transitions (as in $\gamma$ Cas). In 
cycle A, we have no observations previous to the onset of the transitions

\item For the single-peak phase that occurred around October 1996 (last
single peak of cycle B), which is the only phase that we can time with some
accuracy, we see that the symmetry was V$>$R before the single peak
and R$>$V after the single peak, as in $\gamma$ Cas.

\item In cycle B, the single peaked phase close to January 1995 was
accompanied by very bright infrared and optical magnitudes ($J=10.9$ on
Jan 3). The shell
phase during the summer was accompanied by much fainter magnitudes
($J=12.1$ in late July). The infrared magnitudes reached a
maximum close to the next single-peak episode ($J=10.8$ on Oct 15) 
and then faded as
the disc dispersed. Similarly, during cycle A the transition between
a shell phase and a single peak that took place close to Jan 1991, was
accompanied by a brightening in the $I$ band by more than 1 mag (N97).
In contrast, the large increase in H$\alpha$ EW that immediately
preceded the 1992 disc loss, was not accompanied by a brightening of
the source (Unger et al. 1998;N97). 
\end{enumerate}

In summary, insofar as the observations constrain the model, they
favour a close parallel with the spectacular variations
of $\gamma$ Cas. The only point in which the parallel is not obvious are
the large changes in TBW that accompanied the transitions in 
$\gamma$~Cas. As a matter of fact, it is very difficult to estimate
the TBW of the emission lines in V635~Cas 
because 1) the continuum is difficult to
determine (as discussed in Paper I) and 2) even when the overall
profile is a narrow single peak, there is a broad base component, 
(as illustrated in Fig.~\ref{fig:tbw}, but see also the 12th Jan 1996
profile in Fig.~\ref{fig:discloss}) on top of which it sits.

Furthermore, in order to test the validity of the explanation,
we have calculated 
synthetic line profiles for the disc model with the 
parameters listed in Table 1 of Paper I and a viscosity parameter
$\alpha=0.1$ (i.e., the disc whose structure is shown in Fig.~5 of
Paper I) and allowed the inclination
angle $i$ to vary. Since disc truncation will lead to higher densities in the
disc than in isolated Be stars, the line optical depth when the disc
is seen pole-on has been taken to be $\tau = 10^{4}$.
The profiles were computed using 
the same method as described in Okazaki (1996), but now the stellar
continuum and the deviation from the second energy level of hydrogen
have been included. We have assumed that the stellar source function
is equal to the Planck function at $T_{\rm eff}$ and that the
deviation factor, $b_2$, is given by $b_2 = 1/W$, where $W$ is the
dilution factor defined by $W = {1\over2}\{ 1- \left[ 1-(R_{*}/r)^2
\right] ^{1/2}\}$ (e.g., Hirata \& Kogure 1984).

\begin{figure*}[ht]
\resizebox{12cm}{!}{\includegraphics{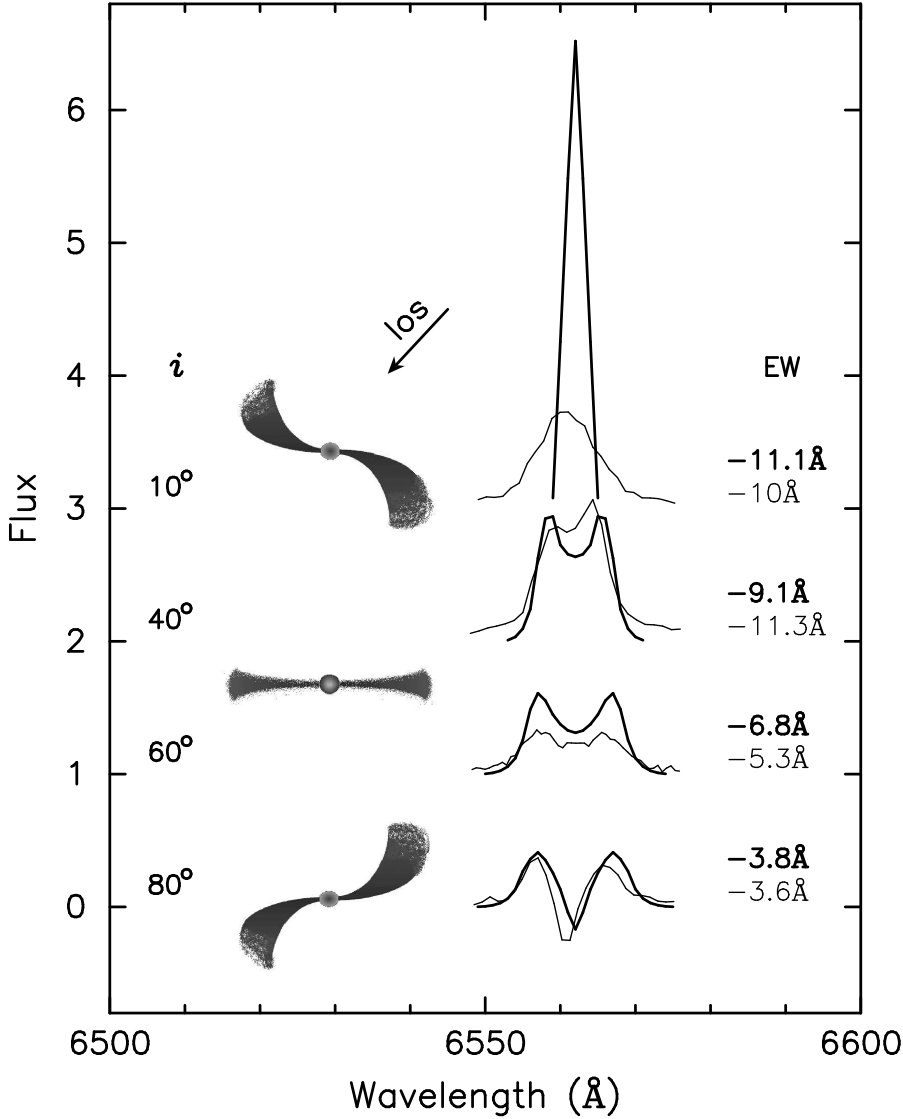}}
\hfill
\parbox[b]{55mm}{
\caption{Synthetic line profiles computed using a disc model with
the parameters for V635 Cas derived in Paper I (thick lines) are 
compared to observed profiles (thin lines). The inclination 
angle $i$ has been allowed to vary in order to reproduce the tilt
caused by disc warping. Annotated on the right hand side of each profile
is the model EW for the line (thick print) and the corresponding
measured EW on the observed profile (thin print).
\newline
\hspace*{15pt} The profiles chosen for comparison are those that
most closely match the theoretical lines, but the actual inclination
angle of the disc under which they were produced (which is, of course,
unknown) may well be different from the inclination chosen for the
model.
\newline
\hspace*{15pt}
A schematic model of the disc configurations that would produce the
profiles is included. The line of sight is supposed to have an
inclination with respect to the equatorial plane of $\sim
50^{\circ}$. The top configuration will produce profiles as
those observed (i.e., with a narrow peak coming out of a broad base),
rather than the simple single-peak theoretical profile, which
corresponds to a purely tilted disc. Due to the disc warp, the central
regions of the disc, which are still seen under a moderate inclination
angle, contribute high-velocity components, which result in the
observed broad base (see Fig.~\ref{fig:tbw}).}
\label{fig:theorylines}}
\end{figure*}

As can be seen in Fig.~\ref{fig:theorylines}, the theoretical profiles
match very well the observed profiles for moderate and high inclinations
(the small difference between the observed and calculated shell profile
being attributable to the V$>$R asymmetry in the former, which was not
considered in the latter), but fail completely to describe the single-peaked
profiles. Our interpretation for this is that the disc is not tilted 
as a solid body but warped (see Fig.~\ref{fig:theorylines}) and therefore the 
central parts of the disc are still seen under a moderate inclination and
contribute high velocity components to the emission lines that create the 
observed broad base (which, as expected, has a TBW similar to that of
double-peaked profiles).

We conclude then that the alignment with the line of sight of 
very different geometrical configurations of a precessing
warped disc is the cause of the shell/single-peaked transitions in 
V635~Cas. We can then divide each of the disc formation and dispersal 
cycles into a number of phases:
\begin{enumerate}
\renewcommand{\theenumi}{(\arabic{enumi})}
\item A disc-less phase, where low level emission can be occasionally
seen on the line wings.
\item A growing-disc phase, where the EW of H$\alpha$ is observed to
grow, while its FWHM decreases (as seen at the start of cycles B and C).
During this phase, there can be fast (days) V/R variability 
(cf. Fig.~\ref{fig:warp}), but
also large-scale V/R variations with longer (many months) quasi-periods,
as during the growing phase of cycle B (cf. Fig.~\ref{fig:vrvar})
\item A warping of the disc, followed by its precession.
It is during this time when the shell/single-peak transitions are seen.
\item The dispersal of the disc.
\end{enumerate}

\section{Global disc oscillations}
\label{sec:GDO}

As mentioned, the emission lines in V635~Cas do generally display
V/R variability. This can be 
in a small scale and rather fast (cf. Fig.~\ref{fig:warp}), and 
probably associated with episodic slightly 
enhanced mass loss from the star (a phenomenon seen in many other Be 
stars with small discs or during disc reformation; cf. Rivinius
et al. 1998), but also on a
much larger scale (as in Fig.~\ref{fig:vrvar}). This second
 variability is similar to the long-term variability
observed in many Be stars and Be/X-ray binaries and can be
attributed to the presence of a $m=1$ mode in the disc, where $m$ is 
the azimuthal wave number (Kato 1983; Okazaki 1991; Papaloizou et al.
1992; Hanuschik et al. 1995; Hummel \& Hanuschik 1997).

Hitherto, the characteristics of the $m=1$ modes in Be discs have been 
studied only for inviscid cases, in which the $m=1$ modes are neutral,
i.e., their frequencies have no imaginary part.
We have studied the applicability of the $m=1$ modes to the disc model
developed for V635~Cas in Paper I. In what follows, we show that the 
$m=1$ modes are overstable in viscous decretion discs. For this purpose, 
we first consider the stability of the local perturbations which vary as 
$\exp [i(\omega t-k_r r-m\phi)]$. After Kato et al.\ (1988) who studied the
stability of viscous accretion discs to pulsational (i.e., axisymmetric)
modes, we can write the local dispersion relation to
non-axisymmetric perturbations as
\begin{eqnarray}
   && (\omega-m\Omega-k_rV_r) \left[ (\omega-m\Omega-k_rV_r)^2
      -\kappa^2-c_{\rm s}^2 k_r^2 \right] \nonumber\\
   && = -2i\alpha \Omega c_{\rm s}^2 k_r^2,
   \label{eqn:disp}
\end{eqnarray}
where $\alpha$ is the Shakura-Sunyaev viscosity parameter,
$c_{\rm s}$ is the sound speed, $V_r$ is the radial component
of the unperturbed velocity field, and $\Omega$ and $\kappa$ 
are the frequency of the disc rotation and the epicyclic
frequency, respectively.

As pointed out by Kato et al. (1988), among three solutions of
Eq.~(\ref{eqn:disp}), it is the inertial-acoustic mode that is
unstable. When the disc is inviscid, the inertial-acoustic mode is
neutral and has the frequency given by
\begin{equation}
   \omega-m\Omega-k_rV_r = \pm (c_{\rm s}^2 k_r^2+\kappa^2)^{1/2}.
   \label{eqn:inviscd}
\end{equation}
However, when the viscous effect is taken into account as a
perturbation, we have
\begin{eqnarray}
   && \omega-m\Omega-k_rV_r \nonumber\\
   && = \pm (c_{\rm s}^2 k_r^2+\kappa^2)^{1/2}
   - i\alpha \Omega {{c_{\rm s}^2 k_r^2} \over
   {c_{\rm s}^2 k_r^2+\kappa^2}}.
   \label{eqn:overstable}
\end{eqnarray}
Hence, the inertial-acoustic mode becomes overstable when the
viscosity is included. The growth rate is given by
\begin{equation}
\left| {\rm Im} \{\omega\} \right| = \alpha \Omega {{c_{\rm s}^2 k_r^2} \over
                {c_{\rm s}^2 k_r^2+\kappa^2}}
              = \alpha \Omega (k_r H)^2,
   \label{eqn:growth}
\end{equation}
where the second equality holds for geometrically-thin, near Keplerian
discs, in which $c_{\rm s} \sim \Omega H$ and $\kappa \sim \Omega$.

\begin{figure}[ht]
\begin{picture}(250,320)
\put(0,0){\includegraphics{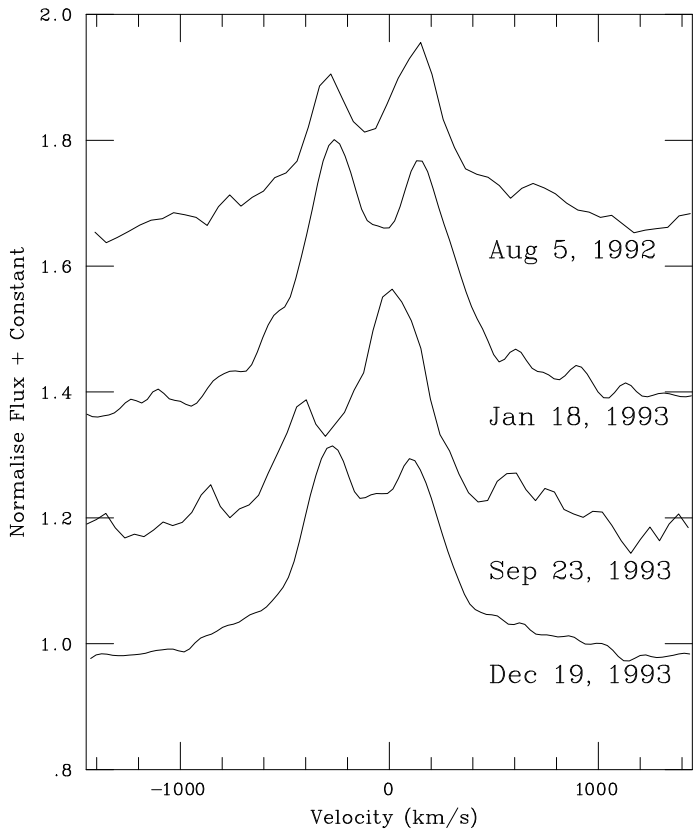}}
\end{picture}
\caption{Evolution of the H$\alpha$ line during 1992\,--\,1993, showing
evidence of a global mode propagating in the circumstellar disc. 
All spectra have been 
divided by a spline fit to the continuum for normalisation and arbitrarily 
offset for display. See Table \ref{tab:halpha} for observational
details of the spectra.}
\label{fig:vrvar}
\end{figure}

Next, we study the stability of global $m=1$ perturbations.
For simplicity, we consider isothermal perturbations.
It can be shown that linearized equations for $m=1$ isothermal
perturbations superposed on the unperturbed axisymmetric 
disc discussed in Paper I are written as
\begin{eqnarray}
  && \left[ i(\omega-\Omega)+V_r{d \over {d r}} \right]
  {\sigma_1 \over \sigma_0}
  +{1 \over {r \sigma_0}}{d \over {d r}} \left( r\sigma_0 v_r \right)
  -{{i v_\phi} \over r} \nonumber\\
  && \hspace*{4em}
  = 0,
  \label{eqn:m1_1} \\
  && c_{\rm s}^2 {d \over {d r}}\left( {\sigma_1 \over \sigma_0}
  \right)
  +\left[ i(\omega-\Omega)+{{d V_r} \over {d r}}+V_r{d \over {d r}}
  \right] v_r - 2\Omega v_\phi \nonumber\\
  && \hspace*{4em}
   = 0,
  \label{eqn:m1_2} \\
  && c_{\rm s}^2 \left(-{i \over r}+\alpha{d \over {d r}}\right)
  {\sigma_1 \over \sigma_0}
  +{\kappa^2 \over {2 \Omega}} v_r \nonumber\\
  && \hspace*{4em}
  +\left[ i(\omega-\Omega)+{V_r \over r}+V_r{d \over {d r}} \right]
  v_\phi = 0,
  \label{eqn:m1_3}
\end{eqnarray}
where $\sigma_0$ and $\sigma_1$ are the unperturbed surface density and
the Eulerian surface-density perturbation, respectively, and $(v_r, v_\phi)$
is the vertically averaged velocity field associated with the
perturbation. As boundary conditions, we impose $(v_r, v_\phi)= {\rm\bf
 0}$ at the inner edge of the disc and $\Delta p=0$ at the outer disc
radius, where $\Delta p$ is the Lagrangian perturbation of pressure.
Solving Eqs.~(\ref{eqn:m1_1})-(\ref{eqn:m1_3}) with the unperturbed
state shown in Fig.~5 of paper I and the above boundary
conditions, we have the fundamental $m=1$ mode shown in
Fig.~\ref{fig:mode}. The period of the mode is 1\,yr and the growth
time is 4.6\,yr.  Note that we have a similar growth time (6\,yr) from
Eq.~(\ref{eqn:growth}) based  on the local analysis, 
taking $k_r \sim \pi/r_{\rm d}$, where $r_{\rm d}$ is
the disc outer radius. Since $r_{\rm d}$ is close to 
the periastron distance in a rough sense, 
 we expect from Eq.~(\ref{eqn:growth}) 
that a system with a larger periastron distance will show a slower 
growth of the $m=1$ mode.

The disc perturbed by the $m=1$ mode becomes eccentric and shows
quasi-periodic changes in Balmer line profiles, the so-called
long-term V/R variations. The slow growth of the $m=1$ mode suggests
that this phenomenon will be observed only after the disc is fully
developed, but before it becomes unstable to warping. During
1992\,--\,1993, 4U~0115+63 exhibited the long-term V/R variations, which
are similar to those of isolated Be stars, except that the period of
variability is much shorter for 4U~0115+63 than for isolated Be stars
(see Fig.~\ref{fig:vrvar}). We note that the characteristics of the $m=1$ mode
shown in Fig.~\ref{fig:mode} are in agreement with these variations. 
The model could also explain the V/R variability observed in 1996\,--\,97, 
though it is also possible that these variations
were due to the warp in the Be disc or, more likely, a 
combination of both effects.

\begin{figure}
\begin{picture}(250,250)
\put(0,0){\includegraphics{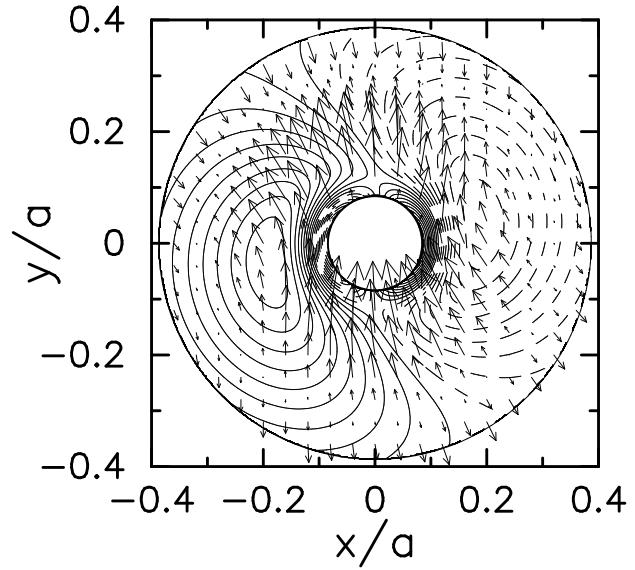}}
\end{picture}
  \caption{Linear, isothermal $m=1$ mode in the viscous disc around
     V635~Cas. The unperturbed disc structure is given
     in Fig.~5 of Paper I. The period of the mode is $1$\,yr
     and the growth time is 4.6\,yr. Both disc and mode rotate
     counterclockwise. The contours denote the relative density
     perturbation in linear scale. The solid (dashed) contours are for
     the region with positive (negative) density enhancement. Arrows
     superposed on the contours denote the perturbed velocity vectors
     normalized by the unperturbed angular velocity, ($v_r/V_\phi$,
     $v_\phi/V_\phi)$.}
   \label{fig:mode}
\end{figure}

\section{Discussion}
\label{sec:discuss}

\subsection{The shell-ejection hypothesis}

Many authors (e.g., Kriss et al. 1983) have attributed the onset of
Type II outbursts in Be/X-ray binaries to hypothetical events of enhanced 
mass loss from the Be star. In this picture, a shell of material is 
ejected from the Be star and, after reaching the orbit of the neutron star, 
part of it is accreted via an accretion disc. 4U\,0115+63 has long been
considered the prototypical system in which this behaviour was observed.
This hypothesis is difficult to hold in view of the observations presented
here, which suggest completely different causes for the observed behaviour.
However, since this is a very extended and generally invoked picture, we
would like to show its lack of consistency, taking as an example
the strong Type II X-ray  outburst
starting on Nov. 18, 1995  (Finger et al. 1995), shortly after the
last single peak phase of cycle B and its associated brightening.

Figure~\ref{fig:bigchange} shows the change in the emission lines that
took place, which must reflect a global and profound change in the  
mass distribution of the disc, since the change in symmetry persisted
for a long time and ultimately led to the dispersal of the circumstellar
disc (see Fig.~\ref{fig:discloss} and the discussion in N98). If we try
to explain all these changes and the Type II X-ray outbursts as 
consequence of the ejection of a shell of material, we find that:
\begin{itemize}
\item Since the shell ejection resulted in a global change of the physical
conditions of the disc, an amount of matter comparable
to a substantial fraction of the disc mass must have been ejected.
\item The shell ejection resulted in a global change of the mass
distribution in the disc, indicating
that it was {\em very} asymmetric. The shell becomes
a blob, with 
$M_{{\rm blob}}> 10^{-10}\:M_{\sun}$ (mass accreted in order
to produce the observed X-ray luminosity).
\item The episode resulted in a global change of the disc structure 
in less than one month. This means that the blob must have been 
ejected with a rather high radial velocity ($\ga 5\,{\rm km\,s}^{-1}$).
\item The episode resulted in a very luminous X-ray outburst -- 
the blob was able to impact the neutron star about one month later.
Such a short timescale suggests that the blob was not circularized and 
merged into the Be disc because of the action of viscous forces.
\end{itemize}

We cannot think of any physical mechanisms within the framework of our
knowledge of Be stars (or stars in general) that could generate such a 
scenario.  We are therefore forced to drop the shell
ejection hypothesis. We note that a modified version of this picture
was defended in N97. This was mainly due to the fact
that, due to sampling effects, shell lines were the most
frequent shape in the sample, leading to the interpretation that they 
represented the quiescence state of the system and therefore that the
system inclination with respect to the line of sight was very large. 

\begin{figure*}[ht]
\begin{picture}(500,280)
\put(0,0){\includegraphics{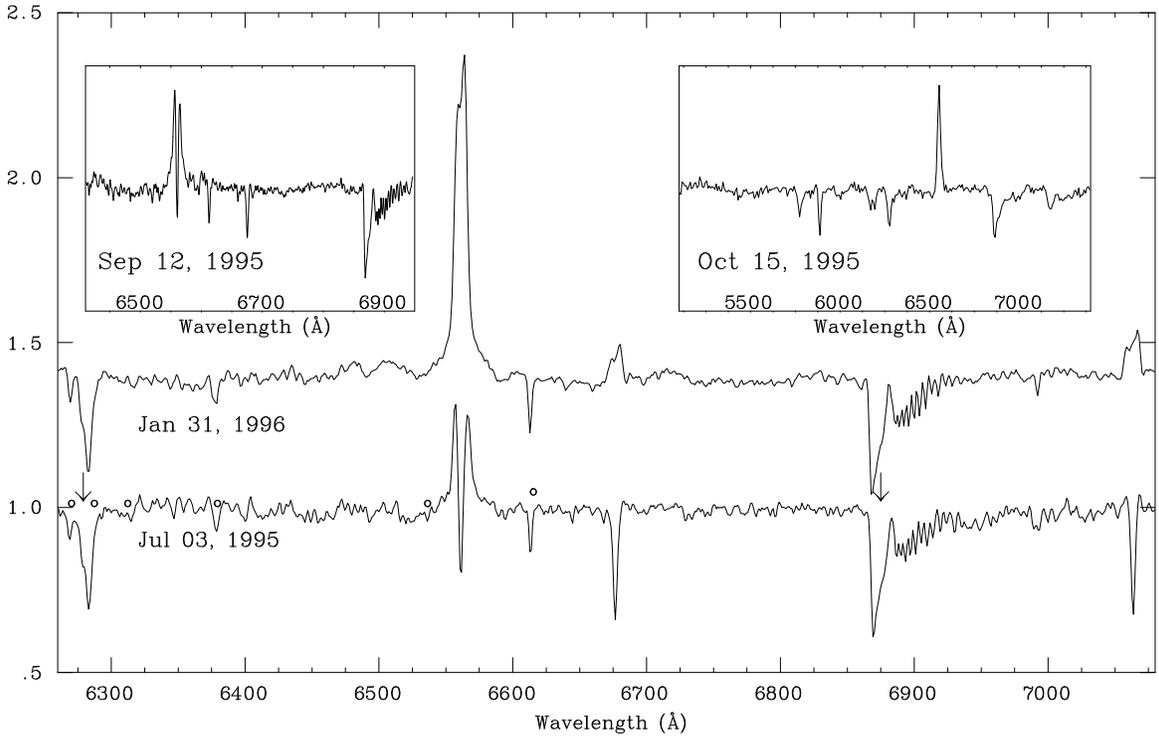}}
\end{picture}
\caption{A large change in the shape of the emission lines took place
immediately before the November 1995 Type II X-ray outburst. The two spectra
in the main panel, taken with the same instrumental setting, allow a direct
comparison of the change in H$\alpha$ and He\,{\sc i} $\lambda \lambda$
6678, 7065 \AA\ lines. The spectra in the two small insets, with
very different resolutions and spectral ranges, strongly constrain the
time during which the change took place. The duration of the event was less
than one month. All spectra have been 
divided by a spline fit to the continuum for normalisation and arbitrarily 
offset for clarity. The strongest diffuse interstellar bands are
marked with a ``$\circ$'' on the spectrum from July 3rd, while the
strongest telluric features are marked by an arrow. See 
Table \ref{tab:halpha} for observational
details of the spectra.}
\label{fig:bigchange}
\end{figure*}

\subsection{The mechanism for Type II outbursts}
\label{sec:quasi}

The body of observations presented here suggests a completely different
explanation for the optical and X-ray behaviour of 4U\,0115+63. The 
dynamical evolution of the (presumably viscous decretion) disc 
surrounding the Be star is responsible for the spectacular changes 
in optical brightness and line shapes, 
and represents the driving force inducing the
X-ray behaviour of the system.

As outlined in Section~\ref{sec:shapes}, the cycles of disc loss and
reformation are the main modulator of the X-ray behaviour. In each
one of those cycles, after the phase of disc growth, the disc
must reach the radius at which it is 
truncated by the presence of the neutron star (presumably, the $4:1$ 
resonance). As a consequence of truncation, the density of the 
circumstellar disc must grow for a time, until the 
disc becomes sufficiently optically thick to continuum
radiation for radiation-driven warping to become possible (see Porter 1998
for a discussion). As the disc warps, the tilt of the outer regions
increases until they come to be close to perpendicular to
the line of sight.

The observation of shell lines alternating
with single-peaked profiles indicates that the disc must
at some point, start precessing.  When the warping is driven by 
radiation, the time-scale of this precession is given by Porter (1998) as
\begin{equation}
T\approx \left( \frac{R_{*}}{R_{\sun}}  \right)^{\frac{5}{2}}
\left( \frac{M_{*}}{M_{\sun}}  \right)^{\frac{1}{2}}
\left( \frac{L_{*}}{L_{\sun}}  \right)^{-1}
{{\rm yr}}
\end{equation}
where we interpret $L_{*}$ as the stellar luminosity to which the
disc becomes optically thick. In the case of
V635~Cas, in which several per cent of the total luminosity is expected 
to contribute to drive the warp, we have $T \approx O(100\,{\rm d})$, 
which is a typical time-scale for the changes in the emission lines.

At present, it is not clear if the tidal warping of the neutron star
will contribute to make the disc precess as a rigid body. Early results 
by Papaloizou \& Terquem (1995) and Larwood (1998) 
seem to indicate that tidal torques when the disc is not coplanar
with the binary orbit will lead to a rigid
body precession of the disc. However, recent results by Lubow \& Ogilvie
(2000) indicate that a disc that extends as far as the $4:1$ resonance
is too big for tidal torques to be effective. For this reason, we will
discuss no further the role of the tidal torque. If the orbit was not
coplanar with the equatorial disc, the tidal torque would be acting on it
from the beginning, but the overall behaviour would not be very different.

The observations of V635~Cas show that during cycles A and
B the X-ray outbursts take place during (or shortly before) the
phase of shell/single-peak
transitions. Though we cannot rule out the possibility that
there is no causal connection between these two phenomena (for
example, if the accumulation of a large amount of material is a
precondition for both disc warping and mass transfer to the neutron
star, none of them will happen at the start of the cycle), the large
perturbation induced on the disc by the combined effects of 
global modes and radiative warping offers a good candidate to the
reason why tidal truncation can be occasionally overcome.
We envisage that, since the disc is heavily distorted, its outer
regions, in which the density is high due to the resonant
truncation, can overflow the critical Roche potential towards the 
gravitational well of the neutron star. We speculate that if 
the disc is heavily elongated in the direction towards periastron, as 
the precession time-scale is long compared to the orbital period, it
can supply its dense outer part onto the neutron star for several
orbital periods. Once matter has flown onto the neutron star, the 
onset and duration of the Type II outburst will depend on the timescales
characteristic of the accretion disc around the neutron star

We note that this picture is supported by the observed quasi-periodicity
of X-ray activity {\em inside} one cycle and the association of X-ray
outbursts with single-peaked lines, which suggests that some particular
disc configuration is necessary in order to have X-ray activity, even 
though this does not imply that such configuration must necessarily 
result in an X-ray outburst. We must emphasize that, in this picture, 
the coincidence of single-peaked profiles and X-ray outbursts is due
only to the system orientation to the line of sight and that some other
phase of the shell/single-peak transition could coincide with mass 
transfer to the neutron star if the orbit was orientated differently.

\subsection{Mass-loss estimates}

In order to estimate the amount of material accreted during a Type II
outburst, we will assume a constant luminosity 
$L_{{\rm x}} \approx 3\times10^{37}\:{\rm erg}\,{\rm s}^{-1}$ during one
month. This is probably an overestimate, but will make up for the 
assumption of $\eta =1$ efficiency in the conversion of gravitational
energy into X-rays. The derived accretion rate on to the neutron star 
$\dot{M}_{{\rm x}} = 1.6\times10^{17}\:{\rm g}\,{\rm s}^{-1} =
2.5\times10^{-9}\:M_{\sun}\,{\rm yr}^{-1}$ means that an amount $\approx
2\times10^{-10}\:M_{\sun}$ is accreted.

Since the disc does not disappear during a Type II outburst, but actually
the strength of the lines does not seem to be too much altered, the original
mass of the disc must be comparable to the amount of material accreted. 
Therefore we can assume $M_{{\rm disc}} \sim 5\times10^{-10}\:M_{\sun}$
at least -- this is, for example, one order of magnitude less than 
the estimates of disc mass in X Persei by Telting et al. (1998). Since 
the Type II outbursts have been observed to occur between 1 and 2 
years after the star is seen in a disc-less state, the mass loss 
rate from the Be star
must be $\dot{M}_{{\rm Be}} \ga 5 \times 10^{-10} M_\odot\,{\rm yr}^{-1}$ 
while the disc is reforming. 

Hanuschik et al. (1993) showed that during mass loss events in
the Be star $\mu$ Cen, the mass loss rate was $\approx 
4\times10^{-9}\:M_{\sun}\,{\rm yr}^{-1}$. This events lasted typically
a few days and the amount of material supplied was not enough to build
a disc. V635~Cas would only need to be supplying a continuous mass loss 
rate one order of magnitude smaller than that from $\mu$ Cen during outbursts
in order to rebuild its disc in the 
$\sim 6$ months determined by our observations. If this material was
reaching the neutron star, it would permanently provide enough material
for an X-ray luminosity 
$L_{{\rm x}} \approx 6 \times 10^{36}\,{\rm erg\,s}^{-1}$, which is 
comparable to that of Type~I X-ray outbursts.

Since 4U~0115+63 is not displaying persistent X-ray emission at such 
level (or any detectable level), it is clear that most of the material lost
from the Be star is kept in the disc and does not reach the vicinity
of the neutron star. Tidal truncation provides an obvious mechanism to
prevent material from leaving the disc during the normal state of the
system. When dynamical instability leads to the disruption of the disc,
material is fed on to the neutron star at a much higher rate than it
is leaving the Be star. This means that no ``enhanced mass loss phase'' 
from the Be star is necessary in order to explain the occurrence of 
X-ray outbursts. 
Moreover, since only a fraction of the disc mass will be fed on to the 
neutron star during a Type II outburst, it is even likely that most
of the disc material will fall back on to the Be star during the 
disc loss process, supporting the idea by Porter (1999) that most of the 
material in a decretion disc must be reaccreted by the Be star.

\subsection{Implications for other Be/X-ray binaries}
  
We have developed a model for the Be/X-ray binary 4U\,0115+63 and used
it in combination with an empirical approach to explain the phenomenology
of this source. The Be/X-ray transient
4U\,0115+63 was selected for careful monitoring because its typical 
timescales seem to be shorter than those of other Be/X-ray binaries, and it
is therefore only natural that our model has been first applied to it. 
However, there is nothing in the derivation of the model or the 
explanations for the phenomenology of 4U\,0115+63 that seems to be exclusive 
to this system, and we are tempted to attribute the speed of the changes
in comparison to similar systems to the closer orbit of the neutron star.

The main conclusions of the application of the viscous decretion model to 
the Be discs in Be/X-ray binaries hold for other systems, i.e., their discs
will be truncated by the resonant interaction of the neutron star. A detailed
analysis of the parameter space and the possible behaviours derived is left
for a forthcoming paper. However it looks very likely from the results of 
this paper that the two types of X-ray outbursts in Be/X-ray transients must
be associated with two different mechanisms or phenomena. 

The explanations
advanced for the behaviour of 4U\,0115+63 further support the picture
developed in N98 in which Type II outbursts are 
associated with catastrophic perturbations of the circumstellar discs.
Even though N98 could not decide in which direction
there was a causal connection between the perturbation and the outburst,
the case of 4U\,0115+63 strongly suggests that
it is the perturbation in the disc which leads to the outburst, and not
otherwise. 
Bildsten et al. (1997) speculated that the Type II outbursts of Be/X-ray
transients could be caused by a thermal instability in an accretion disc
surrounding the neutron star. The results presented here are difficult to
reconcile with such a mechanism, since one
would expect that, in that case, the onset of 
Type II outbursts would be  completely independent of the behaviour of the 
decretion disc surrounding the Be star, in open contradiction with the
observations presented here and in N98.

The large perturbation in the disc of V725 Tau, the optical
counterpart to A\,0535+26, which occurred at the time of its last
Type II outbursts (N98) has led to the gradual
dissipation of the circumstellar envelope (see Haigh et al. 1999;
Negueruela et al. 2000). Such behaviour is reminiscent of that observed in
V635 Cas and is probably indicating some similarly quasi-cyclical
activity on a longer time-scale. Since these two sources are the
Be/X-ray transients for which more regular optical monitoring 
has been presented, it is tempting to generalize this result. Both the 
observations and the estimate of mass loss rates and accretion rates 
during X-ray outbursts support the idea that Type II outbursts in Be/X-ray 
transients are associated with major
dynamical disturbances of the Be star disc rather than 
``shell ejection'' or ``enhanced mass loss''.

However, we do not imply that large perturbations are a necessity for
X-ray outbursts in all
Be/X-ray transients.
As a matter of fact, a whole cycle of of disc growth and dispersal
was observed between 1996 and 2000 in LS~992, the optical counterpart
to  RX\,J0812.4$-$311 (Reig et al. 2000). The time-scales for the cycle, 
for disc loss and disc formation were very similar to those seen in
V635~Cas. However, the X-ray behaviour was completely different. As
soon as the disc reached a certain size, the source started a series
of Type I outbursts which lasted while the disc was present (Reig et al.
2000). We speculate
that in this system (whose orbital parameters are, except for the presumed
81-d orbital period, unknown) the truncation mechanism is not as efficient as
in 4U\,0115+63 and the disc can transfer mass to the neutron star at every
periastron passage. It is even likely that it is precisely because the
efficient truncation of the disc does not allow mass transfer to the
neutron star during long periods (preventing Type I outbursts) that  
the disc in V635~Cas 
experiments large-scale perturbations and  Type II outbursts occur.

Finally, both observational and theoretical results for 4U\,0115+63
confirm the hypothesis advanced for V\,0332+53 (Negueruela et al.
1999) that the long periods of quiescence of Be/X-ray transients can be
due to the limitation of the disc size by the neutron star. 
Therefore we can conclude that the mechanism
controlling the overall X-ray behaviour of the systems is the dynamical 
interaction of the Be decretion disc with the neutron star.

\section{Conclusions}
We have studied the long-term behaviour of the prototypical Be/X-ray 
transient 4U~0115$+$63/V635~Cas. Long-term monitoring
observations indicate that the disc around V635~Cas undergoes
quasi-cyclic formation and dissipation. The time-scale of this
quasi-cycle is compatible with viscous processes. A
quasi-Keplerian viscous decretion disc model predicts a time-scale 
in good agreement with our observations.
          
Our long-term monitoring observations also revealed that
V635~Cas has exhibited all types (double peaks, single peak,
shell, and absorption) of emission line profiles typically seen in
Be stars. We interpret the drastic changes in the line shapes as due 
to warping and/or tilting episodes of the disc. These large changes in 
the disc are associated with Type~II X-ray outbursts, a fact 
difficult to reconcile with models in which the outbursts are caused
by physical processes that affect only the vicinity of the neutron star.

We have shown that viscous decretion discs are overstable against
global one-armed modes. Modelling the disc around V635~Cas as a
viscous decretion disc naturally reproduces the typical time-scales of
V/R variability observed in the system.

Due to tidal truncation, the Be disc acts as a natural reservoir of
material, storing mass lost from the Be star and then feeding it on to
the neutron star at a very high rate after it has become very perturbed.
We are led to conclude that it is the dynamical evolution of
the circumstellar disc under the influence of the radiation pressure
from the central star and the tidal effect of the neutron star companion
what controls the X-ray phenomenology.

\section*{Acknowledgements}

We are very grateful to the PATT and CAT committees for the allocation of
time in their telescopes. This work would not be possible without
the ING service programme. Special thanks to Don Pollacco for his interest
in the monitoring of V635 Cas. The WHT, INT and JKT are operated on the 
island of La 
Palma by the Royal Greenwich Observatory in the Spanish Observatorio
del Roque de Los Muchachos of the Instituto de
Astrof\'{\i}sica de Canarias. This research has made use of the La
Palma Data Archive and of the Simbad data base, operated at CDS,
Strasbourg, France. Special thanks to Dr Eduard Zuiderwijk for his
help with the La Palma data. Thanks are due to all the astronomers 
that have been involved in this monitoring campaign. James
Stevens obtained and reduced the November 1997 optical photometry.
We thank the referee Wolfgang Hummel for his many careful and
constructive remarks, which certainly contributed to improve
the paper.
Data reduction was mainly carried out using the Liverpool
John Moores University {\em Starlink} node, which is funded by PPARC.
During this work, IN has been supported by a PPARC postdoctoral 
fellowship  and later an ESA external fellowship.

\end{document}